\documentclass[preprint,12pt]{elsarticle}

\usepackage{graphicx}
\usepackage{amsmath}
\usepackage{bm}
\usepackage{amssymb}
\usepackage{color}
\usepackage{soul}
\usepackage{xcolor}

\usepackage[utf8]{inputenc}

\journal{Physica A}

\begin{document}

\begin{frontmatter}


\title{Fluctuation theorems in nonextensive statistics}


\author[label1]{Haridas Umpierrez}
\author[label2,label1]{Sergio Davis}

\address[label1]{Departamento de Física, Facultad de Ciencias Exactas, Universidad Andres Bello. Sazié 2212, piso 7, 8370136, Santiago, Chile}
\address[label2]{Comisión Chilena de Energía Nuclear, Casilla 188-D, Santiago, Chile}

\begin{abstract}
Nonextensive statistics is a formalism of statistical mechanics that describes the ocurrence of power-law distributions in
complex systems, particularly the so-called $q$-exponential family of distributions. In this work we present the use of fluctuation theorems
for $q$-canonical ensembles as a powerful tool to readily obtain statistical properties. In particular, we have obtained strong conditions
for the possible values of $q$ depending on the density of states of the system.
\end{abstract}


\end{frontmatter}

\section{Introduction}

Several families of systems in nonequilibrium steady states, including plasmas~\cite{Lima2000, Ourabah2015, Livadiotis2017} cannot be described by the usual canonical ensembles of
statistical mechanics, but instead follow the so-called $q$-canonical distributions $P(\bm x|\beta, q)$, which for a system with Hamiltonian $H(\bm x)$ are given by

\begin{equation}
P(\bm{x}|\beta, q)=\rho(H(\bm x))=\frac{1}{\zeta}\left[1-(1-q)\beta H(\bm{x})\right]_{+}^{\frac{1}{1-q}},
\label{2.2}
\end{equation}
where $q$ is regarded as an additional, free parameter. Tsallis statistics~\cite{Tsallis1988, Tsallis2009} was proposed originally in 1988 and is widely regarded as an
explanation for these $q$-canonical systems, however there are other alternative frameworks such as superstatistics~\cite{Beck2003,Beck2004,Sattin2006,Hanel2011}. Since their
introduction, there has been interest in the properties of these $q$-canonical systems, particularly in the interpretation and possible values of the nonextensive index $q$ ~\cite{Plastino2013,Plastino2017,Plastino2018}. Notably, in the superstatistical framework, the connection has been made between the value of $q$ and the uncertainty of
the superstatistical temperature~\cite{Beck2004}.

\nocite{Briggs2007} 
\nocite{Sivia2006} 

In this work we want to introduce some recent fluctuation identities in statistical mechanics, namely the conjugate variables theorem (CVT)~\cite{Davis2012,Davis2016} into this
problem. Fluctuation identities can be applied to the standard statistical mechanics successfully, recovering a large number of properties related to the expectation values and
the fluctuations of the Hamiltonian and other observables. We show that the use of these theorems can vastly simplify the computations and reveal useful information of systems
in $q$-canonical ensembles.

\section{Results}

For a system with microstates $\bm x \in V$ we consider the definition for the expectation value of a function $f(\bm{x})$ in
the state of knowledge $I$ as
\begin{equation}
\langle f \rangle_{I}=\int_V d\bm{x} f(\bm{x})P(\bm{x}|I).
\label{2.1}
\end{equation}

Now we will consider the case with $f(\bm x)=H(\bm x)$ the Hamiltonian of the system, which we will assume is bounded from below, that is, $H(\bm x) \geq E_0$.
For this Hamiltonian we will denote the density of states by $\Omega(E)$, given by
\begin{equation}
\Omega (E):= \int_V d\bm{x}\delta(H(\bm{x})-E).
\label{2.4}
\end{equation}

\noindent
Motivated by the large class of systems with constant specific heat, let us assume the form
\begin{equation}
\Omega (E)= \Omega_{0} E^{\alpha}
\label{2.5}
\end{equation}
where $\alpha$ is a system-dependent exponent, and we have set (without loss of generality) $E_0=0$. This form not only includes the ideal gas and systems of classical harmonic oscillators, but sometimes has been used to describe more complex systems~\cite{Sanchez2018}.

\noindent
We can now determine the probability density for the energy as
\begin{align}
P(E|\beta, q) & = \Big<\delta(H-E)\Big>_{\beta, q} \nonumber \\
              & = \int_V d\bm{x}\rho(H(\bm{x}))\delta(H(\bm x)-E) \nonumber \\
              & = \rho(E)\Omega(E),
\label{eq_pdf_E}
\end{align}
and in terms of this, define the expectation value for an arbitrary function of energy $g(E)$ as

\begin{equation}
\left\langle g \right\rangle =\int_{0}^\infty dE \Big(\Omega_{0} E^{\alpha}\Big)\left[\frac{1}{\zeta}(1-(1-q)\beta E)_+^{\frac{1}{1-q}}\right] g(E).
\label{2.6}
\end{equation}

Given that the probability density for the energy, $P(E|\beta, q)$ in Eq. \ref{eq_pdf_E} is non-negative and $\Omega(E)$ is always positive, it is clear that
$\rho(E) \geq 0$ and therefore
\begin{equation}
1-(1-q)\beta E\geq0,
\end{equation}
hence, the energy must also be bounded from above, and we have 
\begin{equation}
0 \leq E\leq\frac{1}{(1-q)\beta}.
\label{2.7}
\end{equation}

\noindent
This allows us to define
\begin{equation}
E_{1} := \frac{1}{\beta(1-q)}
\label{eq_E1}
\end{equation}
as the maximum allowed value of energy under given $\beta$ and $q$. Now the expected energy is
\begin{equation}
\left\langle E \right\rangle_{\beta, q, \alpha} = \frac{\Omega_0}{\eta}\int_{0}^{E_1} dE \;E^{\alpha} (1-(1-q)\beta E)^{\frac{1}{1-q}}E,
\label{2.8}
\end{equation}
with the normalization constant $\eta$ given by

\begin{equation}
\eta = \Omega_0 \int_{0}^{E_1}dE E^{\alpha}(1-(1-q)\beta E)^{\frac{1}{1-q}}.
\end{equation}

\noindent
Taking into consideration the upper limit $E_1$ defined in Eq. \ref{eq_E1}, we finally arrive at

\begin{equation}
\left\langle E \right\rangle_{\beta, q, \alpha} =\frac{\alpha+1}{\beta ((\alpha+1)(1-q)+2-q)}.
\label{2.9}
\end{equation}

Another quantity of interest is the microcanonical inverse temperature, defined by

\begin{equation}
\beta_\Omega(E) := \frac{d}{dE}\ln \Omega(E) = \frac{\alpha}{E},
\label{eq_beta_omega}
\end{equation}
whose expected value we can compute in the same manner as before, obtaining
\begin{equation}
\left\langle \beta_{\Omega} \right\rangle_{\beta, q, \alpha} = \beta((1-q)(\alpha+1)+1).
\label{2.11}
\end{equation}

We could, in principle, compute the variances $\langle(\delta E)^{2}\rangle$ and $\langle(\delta \beta_{\Omega})^{2}\rangle$ using the same explicit approach. However, we will take a
look at a more convenient method of calculating the expectation value and variance of a function.

\subsection{The conjugate variables theorem}
\vspace{5mm}
Now, instead of direct integration we will make use of the conjugate variables theorem (CVT)~\cite{Davis2012}, which for a single continuous
random variable $x \in [a, b]$ takes the form
\begin{equation}
\Big<\frac{\partial \omega}{\partial x}\Big>_I + \Big<\omega\frac{\partial}{\partial x}\ln P(x|I)\Big>_I = 0,
\label{eq_cvt}
\end{equation}
where $\omega(x)$ is an arbitrary, differentiable function, and we have assumed that $P(x|I)$ vanishes at the boundaries $x=a$ and $x=b$. The CVT then provides
a family of expectation identities where $\omega$ can, in principle, be chosen suitably. In our case, the energy $E$ is such that $E \in [0, E_1]$ with
$$P(E=0|\beta, q) = P(E=E_1|\beta, q) = 0,$$ so Eq. \ref{eq_cvt} becomes
\begin{align}
\left\langle \frac{\partial \omega}{\partial E} \right\rangle_{\beta, q} & =-\left\langle\omega \frac{\partial}{\partial
E}\ln((1-(1-q)\beta E)^{\frac{1}{1-q}}E^{\alpha})\right\rangle_{\beta,q} \nonumber \\
 & =\left\langle\omega \left[\frac{\beta}{1-(1-q)\beta E}-\frac{\alpha}{E}\right]\right\rangle_{\beta,q}.
\label{2.13}
\end{align}

The second term in the left expectation corresponds to the microcanonical inverse temperature $\beta_\Omega$ (Eq. \ref{eq_beta_omega}), while the first term is the so-called
fundamental inverse temperature~\cite{Davis2019}
\begin{equation}
\beta_{F}(E) := -\frac{d}{dE}\ln\rho(E).
\label{2.14}
\end{equation}

Let us use the choice $\omega(E)=(1-(1-q)\beta E)g(E)$, where $g(E)=E^{m}$ with $m$ an integer. Under this choice, Eq. \ref{2.13} yields the recurrence relation
\begin{equation}
\left\langle E^{m} \right\rangle_{\beta,q}
=\frac{m+\alpha}{\beta (1+(\alpha+1+m)(1-q))}\left\langle E^{m-1} \right\rangle_{\beta,q},
\label{2.15}
\end{equation}
which will let us easily calculate the expectation values of both the energy and the microcanonical inverse temperature. Now as a starting point of the recurrence,
let us choose $m=1$. We have then
\begin{equation}
\left\langle E \right\rangle_{\beta,q}=\frac{1+\alpha}{\beta(1+(\alpha+2)(1-q))},
\label{2.16}
\end{equation}
which is precisely the same result as Eq. \ref{2.9}, although somewhat rearranged. In this way, we see that for the energy, the use of the CVT is a simple and consistent alternative
method to obtain expectation values. The same holds true for the expectation of $\beta_\Omega$ (Eq. \ref{eq_beta_omega}), for which we choose $m=0$ and obtain
\begin{equation}
\left\langle \beta_{\Omega} \right\rangle_{\beta,q}=\beta(1+(\alpha+1)(1-q)).
\label{2.17}
\end{equation}

\noindent
In order to calculate the variance of the energy, we take $m=2$ to obtain
\begin{equation}
\left\langle E^{2} \right\rangle_{\beta,q}=\frac{\alpha+2}{\beta (1+(\alpha+3)(1-q))}\left\langle E \right\rangle_{\beta,q},
\end{equation}
which can be rearranged as
\begin{equation}
\left\langle E^{2} \right\rangle_{\beta,q}= \frac{(\alpha+2)((1-q)(\alpha+2)+1)}{(\alpha+1)((1-q)(\alpha+3)+1)}\left\langle E \right\rangle^{2}_{\beta,q}.
\label{2.18}
\end{equation}

\noindent
With this, the variance for the energy is
\begin{equation}
\left\langle (\delta E)^{2} \right\rangle_{\beta,q}=\left(\frac{(\alpha+2)((1-q)(\alpha+2)+1)}{(\alpha+1)((1-q)(\alpha+3)+1)}-1\right)\left\langle E \right\rangle^{2}_{\beta,q}.
\label{2.19}
\end{equation}

\noindent
In the same manner, for the microcanonical inverse temperature $\beta_\Omega$ we replace $m=-1$ in Eq. \ref{2.15} and obtain
\begin{equation}
\left\langle \beta_{\Omega}^{2} \right\rangle_{\beta,q}=\frac{\alpha}{(\alpha-1)}\beta^{2}((1-q)\alpha+1)((1-q)(\alpha+1)+1),
\label{2.20}
\end{equation}
hence the variance will be given by
\begin{equation}
\left\langle (\delta\beta _{\Omega})^{2}\right\rangle_{\beta,q}=\left(\frac{\alpha}{(\alpha-1)}\frac{((1-q)\alpha+1)}{((1-q)(\alpha+1)+1)}-1\right)\left\langle \beta_{\Omega} \right\rangle^{2}_{\beta,q}.
\label{2.21}
\end{equation}

\noindent
Finally, let us calculate the expectation value and the variance of the fundamental inverse temperature $\beta_F$. For this, let us take $\omega(E)=1$ so that, together with the
definition of both inverse temperature estimators, Eq. \ref{2.13} leads to $$\left\langle \beta_{\Omega} \right\rangle_{\beta,q} = \left\langle \beta_{F} \right\rangle_{\beta,q}.$$
\noindent
For $\langle \beta_F^2 \rangle_{\beta,q}$ let us use the choice $$\omega(E)=\beta_{F}(E)=\frac{\beta}{1-(1-q)\beta E},$$ from which it follows that
\begin{equation}
(1-q)\left\langle \beta_{F}^{2} \right\rangle_{\beta,q} = \left\langle \beta_{F}^{2} \right\rangle_{\beta,q} - \left\langle \beta_{\Omega}\beta_{F} \right\rangle_{\beta,q}.
\label{2.22}
\end{equation}

Here we see that, in order to determine the variance of $\beta_F$ from Eq. \ref{2.22} we need the expectation of $\beta_{\Omega}\cdot \beta_{F}$. For that, let us return to CVT
and choose $\omega(E)=\beta_{\Omega}(E)$. Then, we have
\begin{equation}
\frac{-1}{\alpha} \left\langle \beta_{\Omega}^{2} \right\rangle_{\beta,q} = \left\langle \beta_{\Omega}\beta_{F} \right\rangle_{\beta,q} - \left\langle \beta_{\Omega}^{2} \right\rangle_{\beta,q},
\end{equation}
from which it follows that
\begin{equation}
\left\langle \beta_{\Omega}\beta_{F} \right\rangle_{\beta,q} = \frac{(1-q)\alpha+1}{(1-q)(\alpha+1)+1}\left\langle \beta_{F} \right\rangle^{2}_{\beta,q},
\end{equation}
and finally
\begin{equation}
\left\langle \beta_{F}^{2} \right\rangle_{\beta,q}=\frac{1}{q}\frac{(1-q)\alpha+1}{(1-q)(\alpha+1)+1}\left\langle \beta_{F} \right\rangle^{2}_{\beta,q}.
\label{eq_betaF_2}
\end{equation}

\noindent
The variance of $\beta_F$ can then be obtaining by rewriting Eq. \ref{eq_betaF_2} as
\begin{equation}
\left\langle (\delta\beta_{F})^{2} \right\rangle_{\beta,q}=\left(\frac{1}{q}\frac{(1-q)\alpha+1}{((1-q)(\alpha+1)+1)}-1\right)\left\langle \beta_{F} \right\rangle^{2}_{\beta,q}.
\label{2.23}
\end{equation}

\section{Discussion}
\vspace{5mm}
In order to gain some intuition on the obtained results, let us consider the case of $\bm{x}$ consisting of $n$ quadratic degrees of freedom, for which the density of states
is~\cite{Greiner2012},
\begin{equation}
\Omega(E; V, n) = \Omega_0(V, n)E^{\frac{n}{2}-1},
\end{equation}
hence $\alpha = \frac{n}{2}-1$. For instance, this is the case of an ideal gas of $N$ particles in $D$ dimensions, where $n=N\cdot D$. First, it is straightforward to see
that the canonical ensemble expressions are recovered when $q \rightarrow 1$. That is, Eq. \ref{2.16} becomes

\begin{equation}
\langle E \rangle_{\beta, n} = \frac{n k_{B}T}{2},
\end{equation}
which is the canonical equipartition theorem~\cite{Greiner2012}. Together with this, the expectation of the microcanonical inverse temperature given by Eq. \ref{2.17}
becomes simply
\begin{equation}
\langle \beta_{\Omega} \rangle_{\beta, n} = \beta,
\end{equation}
as it should in the case of a canonical ensemble. This also gives the expectation of the fundamental inverse temperature as $\beta$. The variance of the energy becomes
\begin{equation}
\left\langle(\delta E)^2\right\rangle_{\beta, n} = \frac{n (k_{B} T)^{2}}{2},
\end{equation}
corresponding to the well-known formula connecting energy fluctuations with the heat capacity at a constant volume in the canonical ensemble. From all this it is clear that
all energy-dependent thermodynamic properties are correctly recovered.

An interesting fact arises when we study the variance of the inverse temperature estimators, since in the canonical ensemble, the concept of fluctuations of temperature is unclear,
and actually has not been devoid of controversy~\cite{Kittel1988, Mandelbrot1989}. As $\beta$ is strictly fixed it cannot fluctuate, however the variances of both estimators
$\beta_{\Omega}(E)$ and $\beta_{F}(E)$ exist in this model (because $E$ itself is allowed to fluctuate) and have a well-defined expression for $q=1$. This may seem paradoxical at
first, however there are two ways to approach this apparent contradiction. The first is the traditional interpretation in superstatistics of the variance as a result of
spatio-temporal variations on the temperature, motivated by the fact that it is possible to recover $q$-canonical ensembles from particular assumptions about the microscopic dynamics
of a system~\cite{Sattin2004}. The second approach to this point regards the variance as a product of the uncertainty as to the actual value of the (unique) temperature of the system;
this is compelling when deriving the $q$-canonical ensembles through marginalization and Bayesian probability rules~\cite{Sattin2006,Davis2018}. In such interpretation, the
nonextensive index $q$ is directly related to the lack of information we have about the system.

Examining the variance of the microcanonical inverse temperature $\beta_\Omega$, given in Eq. \ref{2.21}, we can notice a pole at $\alpha = 1$, which corresponds to $n=4$. This behaviour
does not appear in the variance of the fundamental inverse temperature (Eq. \ref{2.23}). Moreover, when taking $q \rightarrow 1$, Eq. \ref{2.21} becomes
\begin{equation}
\left\langle (\delta \beta_{\Omega})^2_{\beta} \right\rangle=\frac{\beta^{2}}{\alpha-1},
\label{2.24}
\end{equation}
which again has a pole at $\alpha=1$ and presents negative values below that, which of course cannot be realized for variances. This is an indicator that the microcanonical
inverse temperature fails to be a reliable estimator for temperature. On the other hand, when taking $q \rightarrow 1$ in Eq. \ref{2.23}, the variance of $\beta_F$ becomes zero
as expected, since in the canonical ensemble the fundamental inverse temperature is precisely the constant $\beta$.

When taking the thermodynamic limit, i.e. when $\alpha \rightarrow \infty$, the variances of both the energy and the microcanonical inverse temperature go to zero. This is to be
expected, as $\beta_\Omega$ becomes a constant when the fluctuations of $E$ vanish. However, the variance of the fundamental inverse temperature in this limit is
\begin{equation}
\left\langle (\delta\beta_{F})^2 \right\rangle_{\beta,q}=\left(\frac{1-q}{q}\right)\left\langle\beta_{F}\right\rangle^{2}_{\beta,q},
\label{2.25}
\end{equation}
which is not manifestly zero. This is counterintuitive since not only $\beta_F$ is a function of the energy (and should have zero variance as is the case with $\beta_\Omega$), but
also because in the thermodynamic limit, the distributions of any parameter will collapse to its observed value. All of this seems to suggest that the only consistent case in the thermodynamic limit is $q=1$. In order to better understand the behavior of this quantity we will expand it as
\begin{equation}
\frac{\left\langle (\delta \beta)^2 \right\rangle_{\beta,q}}{\beta^{2}}=\frac{\alpha^{2}(1-q)^{3}}{q}+\frac{2\alpha(1-q)^{2}}{q}.
\label{2.26}
\end{equation}

The first thing to note is that Eq. \ref{2.26} has been written with the $\beta$ parameter on the left side so that it can be immediately seen as an dimensionless function of
$q$ and $\alpha$. Now, in the simultaneous limit $\alpha \rightarrow \infty$ and $q \rightarrow 1$ there is a tradeoff between the growth rate of $\alpha$ and the vanishing of $(1-q)$,
resulting in that the whole expression vanishes because of the higher powers of $1-q$ in both terms.

Secondly, the fact that this happens in the variance of the fundamental inverse temperature $\beta_F$ suggests that this inverse temperature has information about the ensemble
that the microcanonical inverse temperature $\beta_\Omega$ does not possess. Moreover, the variance of $\beta_F$ is always higher than the variance of $\beta_\Omega$, and so the
idea that the value of $q$ is a measure of the uncertainty we have on any given thermodynamic quantity, reinforces the previous points and again suggests that the fundamental
inverse temperature contains a more accurate depiction of the ensemble.

Another, more fundamental aspect of Eq. \ref{2.23} is that, being a variance, it has to non-negative, and moreover the square of the expectation value of $\beta_{F}$ is positive.
From this it follows that
\begin{equation}
\frac{(1-q)\alpha+1}{q((1-q)(\alpha+1)+1)} \geq 1.
\label{2.28}
\end{equation}

For the range $0<q<1$, Eq. \ref{2.28} does not deliver any new information. However, for $q > 1$ some considerations must be taken into account. The first case is
\begin{equation}
(1-q)\alpha+1<0\qquad \text{and} \qquad q((1-q)(\alpha+1)+1)<0,
\end{equation}
which corresponds to
\begin{equation}
q>1+\frac{1}{\alpha+1}.
\label{2.29}
\end{equation}

\noindent
When taking these conditions, Eq. \ref{2.28} gives
\begin{equation}
    \alpha \leq -1 \qquad \text{and} \qquad (1-q)^{2} \geq 0,
    \label{2.30}
\end{equation}
which however is ruled out, together with the first condition (Eq. \ref{2.29}), because we know the minimum value of $\alpha$ is $-\frac{1}{2}$. Therefore, only the following
case must hold,
\begin{equation}
    (1-q)\alpha+1>0 \qquad \text{and} \qquad q((1-q)(\alpha+1)+1)>0
\end{equation}
from which it follows that
\begin{equation}
q < 1+\frac{1}{\alpha+1},
\label{eq_lutsko}
\end{equation}
precisely the upper bound shown recently by Lutsko and Boon~\cite{Lutsko2011}, depending on the value of $\alpha$. With this inequality, Eq. \ref{2.28} becomes
\begin{equation}
\alpha \geq -1 \qquad \text{and} \qquad (1-q)^{2} \geq 0.
\label{2.32}
\end{equation}

\begin{figure}[h!]
\centering
\includegraphics{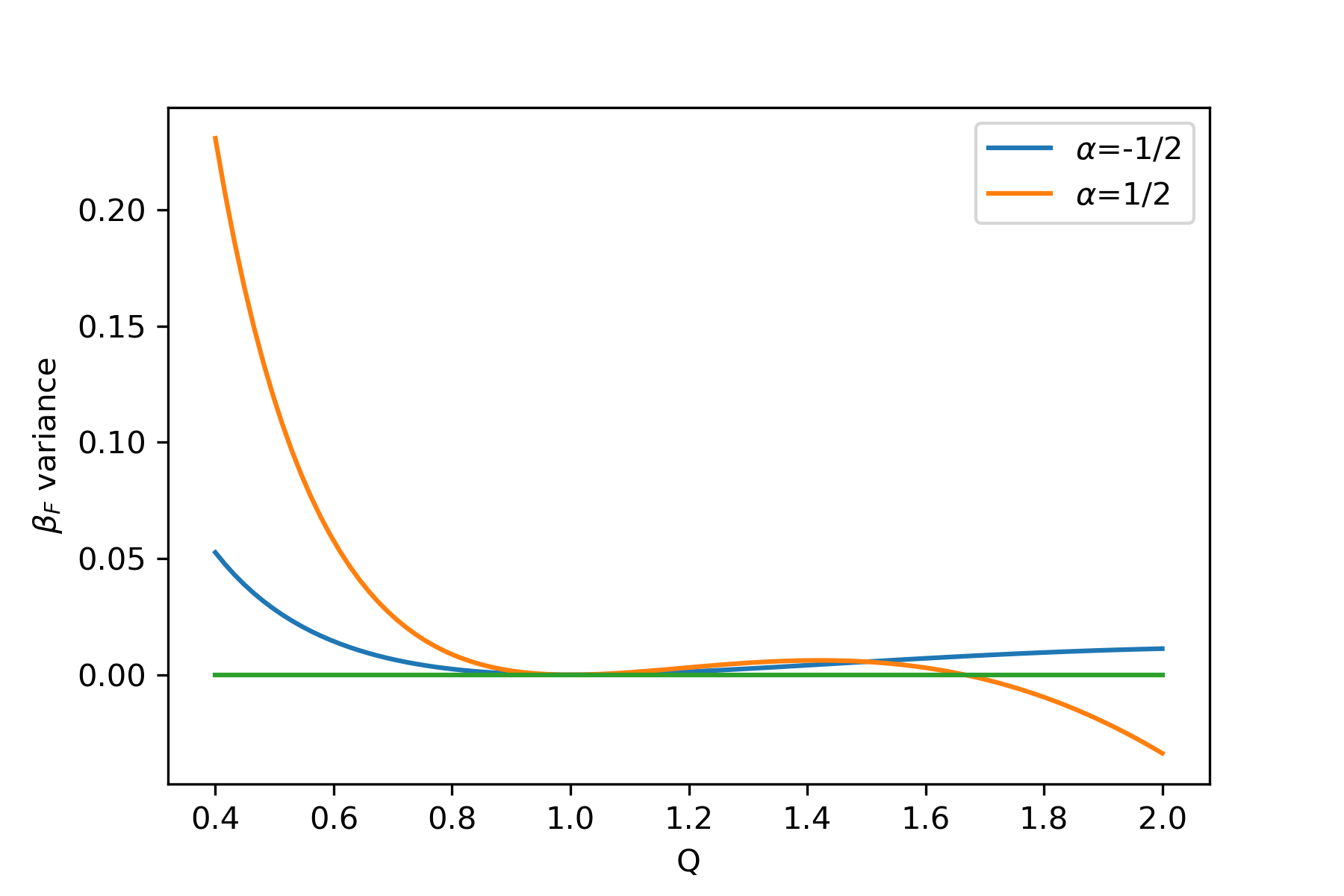}
\caption{Variance of the fundamental inverse temperature $\beta_{F}$ with $\beta=0.3$ as a function of $q$, for different values of $\alpha$.}
\label{Bfvar}
\end{figure}

The characteristic point $q_{LB} = 1 + 1/(\alpha + 1)$ also appears in the plot for the variance of $\beta_{F}$, shown in Fig. \ref{Bfvar}, where from $q_{LB}$ onwards
the variance takes negative values (which is of course not admissible) for any positive $\alpha$.

Having taken a look at the temperature estimators, let us now explore the behavior of the energy. Specifically, we know that
\begin{equation}
0 \leq \frac{\left\langle E \right\rangle_{\beta,q}}{E_{max}}\leq1.
\label{3.1}
\end{equation}

Now, as we know, $E_{max}=\frac{1}{(1-q)\beta}$. Therefore, using this and Eq. \ref{2.16}, the inequality given in Eq. \ref{3.1} becomes
\begin{equation}
0 \leq \frac{(1-q)(1+\alpha)}{(1-q)(2+\alpha)+1}\leq1.
\label{3.2}
\end{equation}

In the range $0<q<1$, the bounded quantity in Eq. \ref{3.2} is always positive for all $\alpha$, so the lower bound of zero holds true. Moreover, imposing the upper limit
in this case gives $q \leq 2$, which is also true. For $q > 1$, given that the bounded ratio in Eq. \ref{3.1} is always positive we have that

\begin{equation}
    (1-q)(2+\alpha)+1 < 0,
    \label{3.3}
\end{equation}
in other words,
\begin{equation}
    q>1+\frac{1}{2+\alpha}.
    \label{3.4}
\end{equation}
With these considerations, Eq. \ref{3.2} imposes that
\begin{equation}
    q \geq 2.
    \label{3.5}
\end{equation}

This condition is always stronger than the one in Eq. \ref{3.4}, and so all values of $q$ between $1$ and $2$ are excluded. However, the condition in Eq. \ref{eq_lutsko} tells us
that $q$ cannot go over $q_{LB}$, and since we have to take them into account simultaneously, $q$ is then allowed to exist in the ranges $0<q\leq 1$ and $2 \leq q < 1+\frac{1}{\alpha+1}$,
the latter only for $\alpha \leq 0$. Because $\alpha \geq -\frac{1}{2}$, the maximum value allowed for $q$ is $3$ in this case, and for sufficiently large values of $\alpha$, $q$ has
no admissible values over $1$.

\section{Concluding remarks}

We have obtained several identities valid for a family of systems in the $q$-canonical ensemble, namely the systems described by densities of states of the form $\Omega(E) \propto
E^{\alpha}$, which is shared by the ideal gas, harmonic oscillators and in general systems with constant heat capacities. Although the systems where $q$-canonical ensembles are observed
may be more complex than this particular form, rather than a depiction of particular systems we aim to explore the use of fluctuation identities such as the conjugate variables theorem as
a tool capable of describing a variety of systems given the form of their density of states. In this sense, the usefulness of this tool to calculate properties of a system is clearly
shown in the analysis done for the different restrictions that arise naturally on the admissible values of $q$. As a quick example, the inequality $q<1+\frac{1}{1+\alpha}$ was obtained by
Lutsko and Boon through an elaborate analysis of the Hamiltonian and the distribution. Here, the same result was obtained in a straightforward manner by examining the variance of the
fundamental inverse temperature $\beta_F$. Furthermore, a second inequality ($q \geq 2$ for $-\frac{1}{2} \leq \alpha \leq 0$) was also unveiled from a condition over the expected energy, which again shows the advantages of working with expectation values of the observables of the system.
Overall, the use of the conjugated variables theorem alongside a model for the density of states proves to be a very efficient way to study systems described by $q$-canonical ensembles.

\section{Acknowledgments}

SD gratefully acknowledges funding from CONICYT Anillo ACT-172101 grant.


\begin{thebibliography}{10}

\bibitem{Lima2000}
J.~Lima, R.~Silva, and J.~Santos.
\newblock Plasma oscillations and nonextensive statistics.
\newblock {\em Phys. Rev. E}, 61:3260--3263, 2000.

\bibitem{Ourabah2015}
K.~Ourabah, L.~A. Gougam, and M.~Tribeche.
\newblock Nonthermal and suprathermal distributions as a consequence of
  superstatistics.
\newblock {\em Phys. Rev. E}, 91:12133, 2015.

\bibitem{Livadiotis2017}
G.~Livadiotis.
\newblock {\em Kappa distributions: Theory and applications in plasmas}.
\newblock Elsevier, 2017.

\bibitem{Tsallis1988}
C.~Tsallis.
\newblock Possible generalization of {B}oltzmann-{G}ibbs statistics.
\newblock {\em J. Stat. Phys.}, 52:479--487, 1988.

\bibitem{Tsallis2009}
C.~Tsallis.
\newblock {\em Introduction to nonextensive statistical mechanics: approaching
  a complex world}.
\newblock Springer Science \& Business Media, 2009.

\bibitem{Beck2003}
C.~Beck and E.G.D. Cohen.
\newblock Superstatistics.
\newblock {\em Phys. A}, 322:267--275, 2003.

\bibitem{Beck2004}
C.~Beck.
\newblock Superstatistics: theory and applications.
\newblock {\em Continuum Mech. Thermodyn.}, 16:293--304, 2004.

\bibitem{Sattin2006}
F.~Sattin.
\newblock Bayesian approach to superstatistics.
\newblock {\em Eur. Phys. J. B}, 49:219--224, 2006.

\bibitem{Hanel2011}
R.~Hanel, S.~Thurner, and M.~Gell-Mann.
\newblock Generalized entropies and the transformation group of
  superstatistics.
\newblock {\em Proc. Nac. Acad. Sci.}, 108:6390--6394, 2011.

\bibitem{Plastino2013}
A.~Plastino and M.~C. Rocca.
\newblock Possible divergences in {T}sallis' thermostatistics.
\newblock {\em EPL}, 104:60003, 2013.

\bibitem{Plastino2017}
A.~Plastino and M.~C. Rocca.
\newblock Analysis of {T}sallis' classical partition function's poles.
\newblock {\em Phys. A}, 487:196--204, 2017.

\bibitem{Plastino2018}
A.~Plastino and M.~C. Rocca.
\newblock Hidden correlations entailed by q-non additivity render the
  q-monoatomic gas highly non trivial.
\newblock {\em Phys. A}, 490:50--58, 2018.

\bibitem{Briggs2007}
K.~Briggs and C.~Beck.
\newblock Modelling train delays with q-exponential functions.
\newblock {\em Phys. A}, 378:498--504, 2007.

\bibitem{Sivia2006}
D.~Sivia and J.~Skilling.
\newblock {\em Data analysis: a Bayesian tutorial}.
\newblock OUP Oxford, 2006.

\bibitem{Davis2012}
S.~Davis and G.~Gutiérrez.
\newblock Conjugate variables in continuous maximum-entropy inference.
\newblock {\em Phys. Rev. E}, 86:051136, 2012.

\bibitem{Davis2016}
S.~Davis and G.~Guti{\'e}rrez.
\newblock Applications of the divergence theorem in {B}ayesian inference and
  {M}axent.
\newblock In {\em AIP Conference Proceedings}, volume 1757, page 020002. AIP
  Publishing, 2016.

\bibitem{Sanchez2018}
E.~Sánchez C. and P.~Vega-Jorquera.
\newblock New {B}ayesian frequency-magnitude distribution model for earthquakes
  applied in {C}hile.
\newblock {\em Phys. A}, 508:305--312, 2018.

\bibitem{Davis2019}
S.~Davis and G.~Gutiérrez.
\newblock Emergence of {T}sallis statistics as a consequence of invariance.
\newblock {\em Phys. A}, 533:122031, 2019.

\bibitem{Greiner2012}
W.~Greiner, L.~Neise, and H.~St{\"o}cker.
\newblock {\em Thermodynamics and statistical mechanics}.
\newblock Springer Science \& Business Media, 2012.

\bibitem{Kittel1988}
C.~Kittel.
\newblock Temperature fluctuation: an oxymoron.
\newblock {\em Phys. Today}, 41:93, 1988.

\bibitem{Mandelbrot1989}
B.~B. Mandelbrot.
\newblock Temperature fluctuations: A well-defined and unavoidable notion.
\newblock {\em Phys. Today}, 42:71--73, 1989.

\bibitem{Sattin2004}
F.~Sattin.
\newblock Superstatistics from a different perspective.
\newblock {\em Phys. A}, 338:437--444, 2004.

\bibitem{Davis2018}
S.~Davis and G.~Gutiérrez.
\newblock Temperature is not an observable in superstatistics.
\newblock {\em Phys. A}, 505:864--870, 2018.

\bibitem{Lutsko2011}
JF~Lutsko and Jean-Pierre Boon.
\newblock Questioning the validity of non-extensive thermodynamics for
  classical hamiltonian systems.
\newblock {\em EPL}, 95(2):20006, 2011.

\end{thebibliography}

\end{document}